# Pressure-Induced Metal-Insulator and Paramagnet-Altermagnet Transitions in Rutile OsO$_2$ Single Crystals

Guojian Zhao[1,6,#], Ziang Meng[1,6,#], Wencheng Huang[2,8,#], Peixin Qin[1,6,*], Shaoheng Ruan[3], Liang Ma[3], Lin Zhu[3], Yuzhou He[3], Li Liu[1,6], Zhiyuan Duan[1,6], Xiaoning Wang[7], Hongyu Chen[1,6], Sixu Jiang[1,6], Jingyu Li[1,6], Xiaoyang Tan[1,6], K. Ozawa[5], Bosen Wang[3], Jinguang Cheng[3], Qinghua Zhang[3,*], Jianfeng Wang[4,*], Chaoyu Chen[2,*], Zhiqi Liu[1,6,9,*]

[1]School of Materials Science and Engineering, Beihang University, Beijing 100191, China
[2]Songshan Lake Materials Laboratory, Dongguan 523808, China
[3]Beijing National Laboratory for Condensed Matter Physics, Institute of Physics, Chinese Academy of Sciences, Beijing 100190, China
[4]School of Physics, Beihang University, Beijing 100191, China
[5]Institute of Materials Structure Science, High Energy Accelerator Research Organization (KEK), Tsukuba, Ibaraki 305-0801, Japan
[6]State Key Laboratory of Tropic Ocean Engineering Materials and Materials Evaluation, Beihang University, Beijing 100191, China
[7]The Analysis & Testing Center, Beihang University, Beijing, 100191, China
[8]Department of Physics and State Key Laboratory of Quantum Functional Materials, Southern University of Science and Technology, Shenzhen 518055, China
[9]Lead contact
*Correspondence: qinpeixin@buaa.edu.cn; zqh@iphy.ac.cn; wangjf06@buaa.edu.cn; chenchaoyu@sslab.org.cn; zhiqi@buaa.edu.cn
#These authors contributed equally to this work.

**ABSTRACT**

Altermagnets with compensated spin structures and nonrelativistic spin splitting have emerged as a new class of magnetic materials. Rutile OsO$_2$ has been theoretically predicted to be altermagnetic, but experimental studies have been limited by synthesis challenges. We have succeeded in synthesizing high-quality single crystals of rutile OsO$_2$. Electrical transport studies reveal that OsO$_2$ is highly conductive and exhibits clear Fermi liquid behavior, indicating strong electron-electron scattering. Magnetic measurements show that the crystals are isotropically paramagnetic. Density-functional theory calculations indicate that bulk OsO$_2$ is semimetallic with coexisting electron and hole pockets, with its magnetic ground state strongly dependent on the on-site Coulomb correlation *U*. Angle-resolved photoemission spectroscopy studies unveil that the bulk bands do not yet show altermagnetic spin splitting. Interestingly, resistivity is rather pressure sensitive: at 44 GPa, a clear metal-insulator transition occurs. Hybrid functional calculations reveal that applying pressure significantly increases the Hubbard *U* value, driving a phase transition from a paramagnetic metal to an altermagnetic metal, and eventually to an altermagnetic insulator. These findings suggest that tuning external pressure effectively modulates the magnetic ground state of OsO$_2$, providing a pathway to realize altermagnetism in this material.

## INTRODUCTION

Unlike ferromagnets and conventional antiferromagnets, altermagnets are an emergent class of magnetically ordered materials characterized by alternating spin polarization in both real and momentum spaces, induced by crystal symmetries. These materials are antiferromagnetically compensated but possess large nonrelativistic spin splitting.[1–3] The rutile metallic oxide $RuO_2$ has been the very first theoretically predicted altermagnet.[4] This pioneering prediction by Šmejkal et al. was based on experimental evidence of antiferromagnetic order in bulk $RuO_2$[5] and thin-film $RuO_2$,[6] which spurred subsequent studies.

Nevertheless, the existence of the long-range magnetic order in bulk $RuO_2$ has been hotly debated.[7–15] On the other hand, the scenario for thin-film $RuO_2$ materials is more complicated due to symmetry breaking at surfaces/interfaces, strain, surface/interface carrier depletion and higher defect densities, all of which can significantly alter its magnetic ground state. Accordingly, a series of experimental results on thin-film $RuO_2$, including the anomalous Hall effect,[16–19] the spin-splitting torque,[20–22] the spin-splitting detected by angle-resolved photoemission spectroscopy (ARPES),[23] the optical second harmonic generation[24] and optical excitation of spin polarization,[25] the inverse spin splitting effect,[26,27] X-ray magnetic linear dichroism,[28] Néel-vector dependent tunneling magnetoresistance[29] and the most recent spin-splitting magnetoresistance effect,[30] are fully consistent with the altermagnetic picture.

$OsO_2$, with the Os element in the same group as the Ru element, closely resembles $RuO_2$, which typically crystallizes in the rutile form (Figure 1A). While $RuO_2$ has been extensively investigated and used in energy applications such as an electrochemical catalyst for oxygen evolution reactions,[31] experimental studies on $OsO_2$ single crystals remain limited[32,33]. This scarcity is partly due to the volatility and toxicity of intermediate $OsO_4$ during single crystal synthesis. What makes $OsO_2$ rather attractive is a recent theoretical report[34] that predicts the altermagnetic behavior parallel to $RuO_2$. In this theoretical study, antiferromagnetism and spin splitting driven by crystal symmetry are forecasted. It is thus an exciting motivation to experimentally synthesize $OsO_2$ single crystals and explore various physical properties.

In this work, we have successfully synthesized high-quality single crystals of rutile $OsO_2$ using a two-step chemical vapor transport method (Figure 1B), overcoming the long-standing challenges associated with its volatile and toxic intermediates. Through comprehensive electrical transport measurements, we demonstrate that $OsO_2$ exhibits excellent metallic conductivity with a clear Fermi-liquid behavior persisting up to 140 K, indicative of strong electron–electron correlations. Magnetic characterization reveals paramagnetic behavior with no evidence of long-range magnetic order. Density functional theory calculations coupled with ARPES further show that the magnetic ground state of $OsO_2$ is highly sensitive to the on-site Coulomb correlation $U$, situating it near a paramagnetic-altermagnetic phase boundary. Moreover, the resistivity of $OsO_2$ exhibits a remarkable metal-insulator transition under a diamond anvil cell pressure of 44 GPa. According to hybrid functional calculations, this applied pressure markedly increases $U$, thereby inducing a sequence of phase transitions, first from a paramagnetic metal to an altermagnetic metal and subsequently to an altermagnetic insulator. Our findings establish external pressure as an effective and direct means to tune the magnetic ground state of $OsO_2$ and induce altermagnetism. A broader perspective suggests that, much like in $RuO_2$-based systems, realizing altermagnetism in $OsO_2$ requires external control of its magnetic order, which can be achieved through diverse approaches including pressure, chemical doping, defect engineering, epitaxial strain, or heterostructuring.

## RESULTS AND DISCUSSION

Figure 1C shows optical images of several $OsO_2$ samples with typical sizes of 0.5–2 mm, which are shiny and golden in color. Almost all the samples possess one or more well-developed flat crystal faces, characteristic of single crystals. X-ray diffraction pattern of a single sample plotted in Figure 1D demonstrates the single-crystalline feature and the major diffraction peaks are fully consistent with the (101) crystallographic plane of rutile $OsO_2$. This indicates that the single crystals of rutile $OsO_2$ have been successfully synthesized. Rietveld refinement (Figure 1E) yields lattice constants $a = b = 4.4955$ Å and $c =$

3.1827 Å. These values agree very well with our DFT-relaxed structure ($a = b$ = 4.522 Å, $c$ = 3.217 Å) [34,35] and previously reported lattice parameters of rutile $OsO_2$, including $a = b$ = 4.4968 Å, $c$ = 3.1820 Å[32] and $a = b$ = 4.500 Å, $c$ = 3.184 Å,[33] confirming the structural consistency and high crystallinity of our samples. Transmission electron microscopy characterizations, (Figure 1F-M) including the selected area electron diffraction, high-resolution atom imaging, electron energy loss spectroscopy, high-angle-annular dark-field imaging, and elemental mapping, further confirm the rutile phase of the sample and indicate the high quality of the single crystal.

Furthermore, X-ray photoelectron and Raman spectroscopies were utilized to analyze the $OsO_2$ single crystal (Figure 1N-Q). The Os $4f_{7/2}$ and $4f_{5/2}$ core-level peaks are observed at binding energies of approximately 51.0 eV and 53.85 eV, respectively, confirming the $Os^{4+}$ valence state. These values are consistent with previous experimental XPS studies[33,36] on $OsO_2$ single crystals, suggesting that the sample is close to stoichiometric composition. We note that the O 1$s$ spectrum also exhibits a higher-binding-energy component at ~532.5 eV, which can be enhanced by surface-sensitive effects such as air exposure and surface cleaning, and is therefore attributed to surface-related oxygen species rather than bulk non-stoichiometry. The Raman shift peaks located at 544, 683, and 725 cm$^{-1}$ correspond to the $E_g$, $A_{1g}$, and $B_{2g}$ modes of rutile $OsO_2$. These values match well with those reported in polarized Raman measurements on $OsO_2$ single crystals, where $E_g$ = 544 cm$^{-1}$, $A_{1g}$ = 685 cm$^{-1}$, and $B_{2g}$ = 726 cm$^{-1}$.[37] Because our Raman spectra were collected without polarization selection, the weak $B_{1g}$ mode near 184 cm$^{-1}$ is not resolved, which is consistent with expectations for unpolarized Raman measurements. All these data corroborate the correct phase of our single crystals.

Figure 2A shows the temperature-dependent linear four-probe resistivity measured with current along the $a$-axis (Crystallographic axes were determined by the selected area electron diffraction in a transmission electron microscopy, based on the microregions of single crystals prepared by focused ion beam milling). As seen from the figure, the room-temperature resistivity of a typical $OsO_2$ single crystal is approximately 30.7 μΩ·cm, which is relatively small for an oxide. This value is lower than that reported in an early study on $OsO_2$ single crystals (~60.0 μΩ·cm),[32] but higher than the room-temperature resistivity of approximately 16 μΩ·cm reported in a previous work.[33] These values are essentially within the same order of magnitude. We therefore attribute the discrepancy in resistivity of $OsO_2$ single crystals from different works primarily to differences in growth conditions, defect scattering and measurement configurations, rather than to a qualitative difference in the intrinsic electronic properties. It is also comparable to the room-temperature resistivity of bulk $RuO_2$ single crystals (~35.2 μΩ·cm).[38] In addition, the metallic transport behavior remains for the entire temperature range from 300 to 2 K, and the residual resistivity ratio (RRR), defined as $\rho$(300 K)/$\rho$(2 K), is approximately 13.2.

It is interesting to notice that the low-temperature resistivity of the $OsO_2$ single crystal exhibits an obvious quadratic temperature dependence. This is a hallmark of Fermi-liquid transport[39] and indicates that electron–electron scattering plays a dominant role in $OsO_2$ at low temperatures. Oxides are strongly correlated electron systems with competing and entangling spin, charge, orbit and lattice degrees of freedom. In some metallic oxides, at low temperatures, the electron-electron scattering could become dominant, leading to the Fermi liquid behavior. For example, electron-doped perovskite $SrTiO_3$,[40] electron-doped cuprates $Nd_{1.85}Ce_{0.15}CuO_{4-x}$,[41] the unusual superconductor $Sr_2RuO_4$,[42] exhibit the Fermi liquid behavior as well.

As shown in Figure 2B, the resistivity below 140 K is plotted as a function of the square of temperature. The clear linear curve demonstrates that the Fermi liquid behavior in the $OsO_2$ single crystal exists in a rather wide temperature range up to 140 K, which is much higher than the upper Fermi liquid temperatures of $SrTiO_{3-x}$ (~80 K)[40] and superconducting $Sr_2RuO_4$ (~30 K).[42] It indicates that the electron-electron scattering in $OsO_2$ is rather strong and is dominant below 140 K, above which the scattering of electrons from phonons surpasses the electron-electron scattering, leading to the deviation of the $T^2$ dependence of resistivity. The fitted linear coefficient for the Fermi-liquid transport is approximately 3.24×10$^{-4}$ μΩ·cm·K$^{-2}$, which is one order of magnitude larger than the Fermi-liquid coefficient of non-strongly correlated metal

Pd[43] (~$3.1 \times 10^{-5}$ $\mu\Omega \cdot$cm$\cdot$K$^{-2}$). Notably, this value is also significantly larger than that reported for bulk RuO$_2$ single crystals (~$5 \times 10^{-5}$ $\mu\Omega \cdot$cm$\cdot$K$^{-2}$),[44] indicating stronger electron–electron correlations and a larger effective quasiparticle mass[45] in OsO$_2$. As the Fermi-liquid coefficient is proportional to the square of the effective mass, the enhanced coefficient reflects stronger correlation effects in OsO$_2$. In addition, even lower temperature resistivity measurements from 2.3 K down to 0.37 K by a Helium-3 module of the physical properties measurement system shows that the resistivity of the OsO$_2$ single crystal is almost constant without any signature of electrical transitions such as superconductivity (Figure 2C).

Subsequent Hall measurements up to 8 T reveal that the predominant carriers are holes at the entire temperature range between 300 and 10 K. As shown in Figure 2D, the room-temperature carrier density is ~$4.45 \times 10^{22}$ cm$^{-3}$. Upon lowering temperature, the carrier density slightly decreases and reaches ~$2.72 \times 10^{22}$ cm$^{-3}$ at 10 K. This value is higher than the carrier density reported for bulk RuO$_2$ single crystals (~$8.5 \times 10^{21}$ cm$^{-3}$ at 10 K),[15] further highlighting differences in electronic structure and transport regimes between the two rutile oxides. The absence of an exponential decrease of carrier density upon cooling down, *i.e.*, the carrier freeze-out effect, implies that the carriers are not majorly from thermally activated defect dopings.[46] Thus, the high hole carrier density on the order of $10^{22}$ cm$^{-3}$ shall originate from the substantial hole pocket density of states at the Fermi level.

The corresponding mobility $\mu$ as a function of temperature is illustrated in Figure 2E. The room-temperature mobility is ~4.57 cm$^2 \cdot$V$^{-1} \cdot$s$^{-1}$, which is similar to the room-temperature mobility of SrTiO$_3$-based Ti 3*d* electron systems.[40,47,48] The mobility increases with decreasing temperature, which is consistent with reduced phonon scattering and is a common trend in metallic and conductive oxide systems. The mobility reaches approximately 92.83 cm$^2 \cdot$V$^{-1} \cdot$s$^{-1}$ at 10 K. This value is much smaller than that reported for bulk RuO$_2$ single crystals, which exhibit a mobility of approximately 5500 cm$^2 \cdot$V$^{-1} \cdot$s$^{-1}$ at 10 K,[15] consistent with the larger carrier density and stronger electron–electron scattering in OsO$_2$, as reflected by its enhanced Fermi-liquid coefficient. It is much smaller than the high mobility (> 10,000 cm$^2 \cdot$V$^{-1} \cdot$s$^{-1}$) of SrTiO$_3$-based electron systems,[40,47,49,50] which may originate from the large carrier density and the strong electron-electron scattering at low temperature evidenced by the Fermi liquid transport behavior.

Out-of-plane magnetoresistance (MR) ranging from 10 to 300 K was systematically measured up to 8 T (Figure 2F). The MR is positive and remarkable only below 100 K. At 8 T, it is ~120% at 10 K and gradually decreases to ~8% at 100 K. This could be understood by the orbital scattering due to the Lorentz force. A prominent feature of this longitudinal magnetotransport is the linear MR effect for all the temperatures below 100 K. A linear MR is useful for magnetic field sensor applications, but the intrinsic mechanisms can be complicated including quantum and classical understandings[40,51] as discussed for nonmagnetic silver chalcogenides,[52] semimetals,[53] and density wave materials.[54] In the case of OsO$_2$, the strong temperature dependence of the magnetoresistance closely follows that of the carrier mobility, suggesting that classical orbital effects associated with the Lorentz force play an important role.[39] Above 100 K, the mobility is substantially reduced, consistent with enhanced inelastic scattering at elevated temperatures (for example, from phonons), and the resulting lower mobility strongly suppresses the orbital magnetoresistance.[16,40,55]

As a long-range antiferromagnetic order is a prerequisite for the emerging altermagnetism, after addressing all the above exotic electrical transport properties, we moved our focus to another pivotal aspect – magnetic properties. The magnetic moment versus temperature for an OsO$_2$ single crystal was measured along different crystallographic directions. Both field-cooling and zero-field-cooling measurements were conducted. As seen in Figure 2G, the magnetization is isotropic along different directions. More importantly, it shows the typical Curie-Weiss behavior, suggesting the paramagnetism. In addition, there is negligible difference between field-cooling and zero-field-cooling measurements, which is consistent with the paramagnetic frame. The following magnetic-field-dependent magnetization measurements result in linear curves as shown in Figure 2H, further supporting the paramagnetism of the OsO$_2$ single crystal.

To rationalize the absence of long-range magnetic order and to assess the proximity of OsO$_2$ to a theoretically proposed altermagnetic state, we performed spin-polarized DFT calculations with and without spin-orbit coupling (SOC), focusing on the bulk band structure and systematically examining the effects of the on-site Coulomb interaction $U$. As shown in Figure 3A-F, a small $U$ of 0 and 1 eV gives rise to a nonmagnetic state, while a large $U$ of 2 eV results in an altermagnetic phase, which is consistent with the scenario of $RuO_2$.[4,16] Also similar to that of $RuO_2$,[4,16] the band of the altermagnetic phase has a large spin splitting along ΓM and ZA lines, where multiple nodal-line band crossings occur in the Brillouin zone. With the larger SOC effect compared to $RuO_2$, the band crossings or degeneracies generate a large local gap. In addition, all the band structures under different $U$ values and SOC situations show a semimetallic scenario with both electron and hole pockets at the Fermi level, which validates the large hole density we have experimentally determined.

To benchmark the electronic structure and constrain the DFT description, particularly the effects of SOC and the on-site Coulomb interaction $U$, we performed ARPES measurement on single crystals of $OsO_2$. In line with the single-crystalline X-ray diffraction results, the cleaved surface corresponds to the (101) crystallographic plane of rutile $OsO_2$. Such a surface BZ can be viewed as the projection of the $\Gamma ZRX$ plane along the [101] direction, with the effective high-symmetry path $\bar{\Gamma} - \bar{X} \sim 0.7$ Å$^{-1}$ and $\bar{\Gamma} - \bar{Z} \sim 0.57$ Å$^{-1}$. Indeed, ARPES Fermi surface mapping shown in Figure 3G follows such periodicity, manifested by the alternating strong intensity at $\bar{\Gamma}$ and weak intensity at $\bar{R}$. To further substantiate the ARPES interpretation, we compare the experimental spectra with DFT calculations performed with and without SOC. As summarized in Figure S2, the SOC-included calculations provide a more faithful description of the low-energy dispersions observed in ARPES, whereas the non-SOC results show noticeable deviations, particularly near the Fermi level. This comparison indicates that SOC is an essential ingredient for capturing the electronic structure of $OsO_2$ and should be included in the ARPES–DFT correspondence.

Although not sharp, one can still resolve at least one elliptic pocket at the extended surface BZ center. Our DFT simulated Fermi surface of bulk band projected to the (101) surface (superposed in the first surface BZ of Figure 3G) not only agrees well with the measurements, but also shows more detailed features. From the ARPES spectral cut along high symmetry paths shown in Figure 3G and 3H, this pocket comes from one electron-type, parabolic band centered at $\bar{\Gamma}$, with its band bottom located around at ~ -0.5 to -0.7 eV and Fermi vector around $\pm 0.4$ ($\pm 0.3$) Å$^{-1}$ along $k_y$ ($k_x$) direction. From the DFT calculation shown in Figure 3A-F, one can find the corresponding electron-type band centered at Γ. As its band bottom shifts towards the Fermi level with increasing $U$ (Figure 3D-F), the identification of such band from ARPES helps us to choose the practical value of $U$ for the DFT simulation. As plotted in Figure 3H and 3I on top of the ARPES spectra, DFT calculated band projections onto the (101) surface with $U$ = 1.0 eV for the non-magnetic state can reproduce the experiments very well. We further check the robustness of this comparison against the on-site Coulomb interaction $U$. As shown in Figure S3, the overall Fermi-surface topology and the low-energy dispersions remain consistent with the ARPES data for $U$ values around 1.0–1.2 eV, while increasing $U$ progressively enhances the band splitting and eventually stabilizes an altermagnetic ground state. Therefore, we adopt $U$ = 1 eV for the comparison, and the results indicate that the studied $OsO_2$ single crystals, although not altermagnetic, lie close to the paramagnetic–altermagnetic phase boundary.

Given that our DFT+ARPES analysis places $OsO_2$ close to the paramagnetic–altermagnetic phase boundary, pressure/strain provides a natural knob to tune its electronic and magnetic ground state.[35,56–58] Recent theoretical calculations have demonstrated that strain can act as an effective tuning parameter to stabilize an altermagnetic phase in $OsO_2$.[35] As $OsO_2$ single crystals have been rarely studied owing to the difficulties in their synthesis, it is thus also rather intriguing for us to explore the electrical transport properties of $OsO_2$ single crystals under different diamond anvil cell (DAC) pressures. For such an experiment, a thin flake sample with a few μm thickness obtained by crushing a mm-scale single crystal was adopted for building linear four-probe electrical contacts as shown in Figure 4A. Temperature-dependent resistance curves are shown in Figure 4B for different DAC pressures. Overall, the resistance of the $OsO_2$ flake is sensitive to DAC pressures and enhanced by increasing the applied pressures. From 10 to 44 GPa, the room-temperature resistance is elevated by more than three orders of magnitude. Starting from 44 GPa, an intriguing pressure-induced metal-insulator transition occurs. Pressure applied in a diamond anvil cell is expected to modify interatomic distances, orbital overlap and correlation strength of electrons. Such a pressure-sensitive characteristic could indeed facilitate the construction of electric field/piezoelectric strain controlled electronic devices based on $OsO_2$.[56,57,59–64] Although our diamond-anvil-cell experiments probe electrical transport rather than magnetic responses, the strong pressure dependence of the resistivity underscores the pronounced sensitivity of $OsO_2$ to lattice perturbations. These results motivate future strain-engineering and pressure-dependent magnetic or spectroscopic studies to explore the theoretically proposed altermagnetic regime.

The potential exotic altermagnetism as predicted for $OsO_2$ from theoretical calculations was based on the predicted antiferromagnetic order. From a theoretical point of view, such an antiferromagnetic order could be obtained by tuning the on-site Coulomb repulsion potential $U$,[34] which is also consistent with our theoretical calculations. As shown in Figure 4C, our calculations demonstrate that the altermagnetism of bulk $OsO_2$ can be stabilized by a condition of $U > 1.2$ eV. The corresponding altermagnetism benefiting from the low crystal symmetry of the simple rutile structure was thus obtained. Although unfortunately the antiferromagnetic order has not been detected in our single crystals, it is still promising to realize altermagnetism in $OsO_2$ once the magnetic order could be established, for instance, via external pressure, and other approaches similar to the $RuO_2$ case.

To gain microscopic insight into the pressure-induced transport evolution and the metal–insulator transition observed in our DAC measurements, we performed high-pressure electronic-structure calculations. The spin-polarized band structures of bulk $OsO_2$ at 50 GPa, calculated for different Hubbard $U$ values, are presented in Figure 4D-I. The bands obtained with $U = 1$ eV and $U = 2$ eV (Figure 4D and 4E) exhibit broader dispersion compared to those at ambient pressure (Figure 3), which aligns with the enhanced orbital overlap expected under compression. Notably, an altermagnetic splitting is consistently observed for $U > 1$ eV (Figure 4E-I). As the $U$ value increases, the bandwidth narrows, leading to a gradual separation between the conduction and valence bands. A metal-insulator transition occurs within the $U$ range of 3–3.5 eV (Figure 4F and 4G). Our electrical resistivity measurements indicate that bulk $OsO_2$ has become an insulating state at 50 GPa (Figure 4B). Hybrid functional calculations predict an insulating phase with a direct band gap of 0.8 eV (Figure 4I), which is consistent with the results obtained from PBE+$U$ calculations with $U = 3.8$ eV (Figure 4H). These computational findings collectively indicate that applied pressure enhances electronic correlations, driving the transition of bulk $OsO_2$ from a metal to an insulator together with a transition from a nonmagnetic to an altermagnetic state.

**CONCLUSIONS**

To summarize, as spurred by the very recent theoretical prediction of a possible altermagnetic phase of $OsO_2$, we have developed an effective two-step approach to successful synthesis of high-quality $OsO_2$ single crystals, which have been extremely challenging to fabricate and hence scarcely investigated. There are multiple intriguing features unveiled for $OsO_2$ in this study: (1) The Fermi liquid behavior persisting up to 140 K implies strong electron-electron scattering; (2) The superior metallic electrical conductivity better than $RuO_2$ single crystals could be promising for establishing an excellent electrochemical catalyst for diverse energy applications; (3) The mechanism of the striking low-temperature linear MR remains elusive but could be useful for sensor applications; (4) $OsO_2$ is semimetallic with both electron and hole pockets at the Fermi level and its magnetic ground state strongly relies on the Coulomb correlation $U$; (5) The pressure-sensitive electrical transport properties can enable the future low-power electronic device applications once $OsO_2$ is integrated with piezoelectric/ferroelectric materials. Regarding the initial motivation – altermagnetism, although the long-range antiferromagnetic order and the altermagnetic spin-splitting have been absent in our bulk $OsO_2$ single crystals, the vital key to realizing that might be artificially engineering its magnetic ground states to accomplish the long-range antiferromagnetic order by experimentally tuning the Coulomb correlation $U$. In this context, we demonstrate a clear pressure-induced metal–insulator transition in $OsO_2$ single crystals. Besides, our high-pressure electronic-structure calculations reveal that external pressure substantially enhances $U$, driving a sequence of magnetic and electronic phase transitions from a paramagnetic metal to an altermagnetic metal, and ultimately to an altermagnetic insulator.[35] Together, these findings establish pressure as a key physical control parameter for the magnetic ground state of $OsO_2$ and provide fundamental insight into the emergence of altermagnetism in correlated oxides.

## METHODS

### Materials and methods

$OsO_2$ single crystals were synthesized via a general two-step approach (Figure 1B). Firstly, 2 g Os powder with 99.8% purity was uniformed mixed with 1.06 g $NaBrO_3$ powder with 99.5% purity. Such a mass ratio was inferred from the molar ratio of Os and O for $OsO_2$ and $NaBrO_3$ was a strong oxidant. The mixed powder was then vacuum-sealed into a quartz tube with a $10^{-5}$ Torr based pressure. Subsequently, the vacuum-sealed quartz tube was placed into a tube furnace. It was firstly kept at 300 °C for 2 days of pre-heating and then heated up to 650 °C for chemical reaction. The chemical reaction time was set as 2 days. After that, the tube was cooled down to room temperature. For any temperature variation process, the ramping rate was kept as 1 °C/min. In the end of this step, polycrystalline $OsO_2$ powder was accumulated in one end of the quartz tube.

Afterwards, the second step was started and the vacuum-sealed quartz tube which had gone through the above first step was further loaded into a double-zone tube furnace. The quartz tube's end with the formed polycrystalline $OsO_2$ powder was kept at 950 °C and the other end was kept at 875 °C. These temperatures were remained for 14 days to achieve chemical-vapor-transported $OsO_2$ single crystals. After that, both ends of the quartz tube were cooled down to room temperature. During this step, the ramping rate for heating up was 50 °C/h below 800 °C and 10 °C/h above 800 °C. For cooling down, the ramping rate was 1 °C/min for both ends.

### Characterization Techniques

### X-ray diffraction

X-ray diffraction measurements were performed via a Bruker D8 Advance diffractometer with Cu-$K\alpha$ radiation ($\lambda$ = 1.54184 Å).

### Transmission electron microscopy

Transmission electron microscopy measurements were carried out using a spherical aberration-corrected JEOL JEM NeoARM200 microscope operated at an acceleration voltage of 200 kV. High-angle annular dark-field STEM images were acquired with collection angles of 75–310 mrad. Electron energy-loss spectroscopy (EELS) measurements were performed with a collection semi-angle of 111 mrad. EELS spectra were acquired using a GIF Continuum spectrometer equipped with a K3 direct electron detection camera.

### X-ray photoelectron spectroscopy

X-ray photoelectron spectroscopy (XPS) measurements were performed at room temperature using a Thermo Fisher Scientific ESCALAB Xi+ system equipped with a monochromated Al $K\alpha$ X-ray source. The beam spot size was approximately 600 μm, and the measurements were carried out in a normal emission geometry. The binding energies were calibrated using the C 1*s* peak at 284.8 eV.

### Raman spectroscopy

Raman spectra were collected at room temperature using a confocal micro-Raman system with a 532 nm laser excitation and the laser power was 4 mW. The measurements were performed in a non-polarized configuration using a 50× VIS LWD objective and a 600 lines/mm grating. The laser was focused onto the sample surface with a spot size of approximately 1–2 μm.

### Electrical and magnetic measurements

For electrical transport measurements, electrical contacts were constructed by an Al wire bonder and the measurements were implemented by a physical properties measurement system. Magnetic moments were measured by a Quantum Design VersaLab system with a vibrating sample magnetometer option. The measurement geometries for electrical transport, magnetoresistance, Hall effect, and magnetization are schematically illustrated in Figure S1.

**High pressure transport measurements**

The high-pressure transport measurements were carried out in a He-4 cryostat using a diamond anvil cell equipped with 200 μm culet diamonds and KBr as the pressure-transmitting medium.

**DFT calculations**

The density-functional theory (DFT) calculations were performed using the Vienna *ab initio* simulation package[65] within the projector augmented wave method[66] and the generalized gradient approximation of the Perdew-Burke-Ernzerhof[67] exchange-correlation functional. The cutoff energy of 520 eV and 12 × 12 × 18 Γ-centered *k*-point meshes were used. The crystal structure of $OsO_2$ was fully relaxed until the residual force of each atom is less than 0.005 eV/Å. The optimal lattice constants under ambient pressure are *a* = 4.522 Å and *c* = 3.217 Å. The SOC effect was considered in our electronic structure calculations. The on-site Coulomb correlation of Os-*d* orbitals within the DFT+*U* approach[68] was used with the effective Hubbard parameter *U* tested, where *U* = 1.0 eV with the non-magnetic state can reproduce the experimental measurements. To simulate the pressure-induced metal-insulator transition, hybrid functional calculations[69] were also employed. The tight-binding Hamiltonians based on the maximally localized Wannier functions[70] with Os-*d* and O-*p* orbitals are constructed to further calculate Fermi surface and surface projection of bulk band by using the WannierTools package.[71]

**ARPES measurements**

ARPES measurements were performed at BL-28A of the Photon Factory (PF), KEK, Japan. We used linearly polarized light of 100 eV. The energy resolution was better than 30 meV, and the angular resolution was better than 0.1°. Temperature during ARPES measurements was set at 12 K. Samples were in situ cleaved along the (101) plane within the tetragonal structure. The detector slit is parallel to the ARPES measurement plane as defined by the incident beam and the normal line of sample surface. The beam spot size is smaller than 15 μm × 15 μm.

**RESOURCE AVAILABILITY**

**Lead contact**
- Requests for further information and resources should be directed to and will be fulfilled by the lead contact, Zhiqi Liu (zhiqi@buaa.edu.cn).

**Materials availability**
- Materials generated in this study will be made available on request.

**Data and code availability**
- Any additional information required to reanalyze the data reported in this paper is available from the lead contact upon request.

**ACKNOWLEDGMENTS**

P.Q. acknowledges the financial support from the National Natural Science Foundation of China (no. 52401300). Z.M. acknowledges the financial support from the National Natural Science Foundation of China (no. 524B2003). L.L. acknowledges the financial support from the National Natural Science Foundation of China (no. 525B2008). Z.L. acknowledges financial supports from the National Natural Science Foundation of China (nos. 52425106, 52121001 and 52271235), the National Key R&D Program of China (nos 2022YFA1602700 and 2022YFB3506000) and the Beijing Natural Science Foundation (no. JQ23005). C.C. acknowledges financial supports from the National Key R&D Program of China (no. 2025YFA1411200) and the National Natural Science

Foundation of China (no. 12574068). J.W. acknowledges financial supports from the National Natural Science Foundation of China (no. 12474217) and the Beijing Natural Science Foundation (no. 1242023). Q.Z. acknowledges financial supports from the National Natural Science Foundation of China (no. 52322212) and the National Key R&D Program of China (nos. 2023YFA1406300 and 2024YFA1409500). J.C. acknowledges financial supports from the National Natural Science Foundation of China (no. 12025408) and the National Key R&D Program of China (no. 2023YFA1406100). This work is supported by National Natural Science Foundation of China (no. U25A20244). This work is supported by the Fundamental Research Funds for the Central Universities. The authors acknowledge the Analysis & Testing Center of Beihang University for the assistance with Helium-3 low-temperature measurements. The authors acknowledge the high-pressure experiments were carried out at the CAC station of Synergetic Extreme Condition User Facility (SECUF) [https://cstr.cn/31123.02.SECUF]. The authors acknowledge Mr. Yehua Huang & Dr. Huiyang Gou at Center for High Pressure Science and Technology Advanced Research (HPSTAR), Beijing for preliminary diamond anvil cell pressure measurements with the van der Pauw geometry.

**AUTHOR CONTRIBUTIONS**

Conceptualization: Z.L., G.Z., Z.M and P.Q.; Methodology: G.Z, P.Q., Z.M. and W.H.; Formal analysis: G.Z., Z.M., P.Q., J.W., C.C. and Z.L.; Investigation: G.Z., Z.M. and W.H.; Data curation: G.Z., Z.M., W.H., P.Q., S.R., L.M., L.Z., S.J., J.L., X.T., K.O., B.W., J.C. and Y.H.; Visualization: G.Z., C.C., J.W., L.L., D.Z., X.W. and H.C.; Supervision: Z.L., C.C. and J.W.; Writing—original draft: G.Z., Z.M. and P.Q.; Writing—review & editing: Z.L., C.C. and J.W.; Z.L. led and coordinated the project.

**DECLARATION OF INTERESTS**

The authors declare no competing interests.

**SUPPLEMENTAL INFORMATION**

**Document S1. Figures S1–S3**

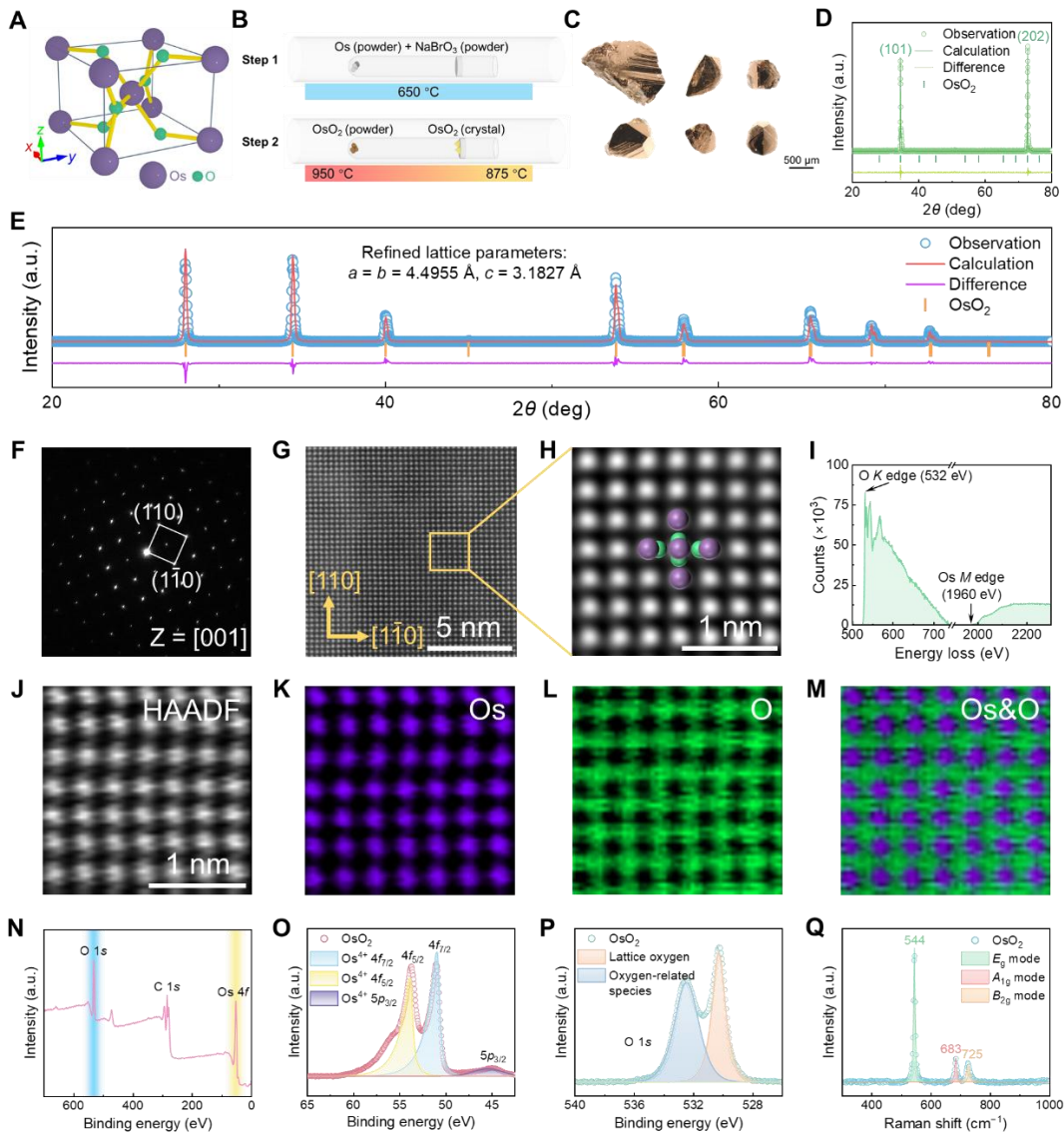

**Figure 1. Crystal structure, synthesis process, and comprehensive structural characterization of $OsO_2$ single crystals.**
(A) Crystal structure of $OsO_2$, where purple and green spheres represent Os and O atoms, respectively. (B) Two-step synthesis of $OsO_2$ single crystals: Step 1, Os powder reacts with $NaBrO_3$ at 650 °C to produce $OsO_2$ powder; Step 2, $OsO_2$ crystals are grown in a temperature gradient from 950 °C to 875 °C. (C) Optical images of representative $OsO_2$ single crystals. (D) XRD pattern collected from a single $OsO_2$ crystal. (E) Rietveld refinement of the XRD pattern of the $OsO_2$ powder obtained by grinding $OsO_2$ single crystals. (F) SAED pattern of $OsO_2$ single crystal along the [001] zone axis. (G) HAADF-STEM image taken along the [001] zone axis. (H) Magnified HAADF-STEM image corresponding to the boxed region in (G), overlaid with the atomic structure of $OsO_2$. (I) EELS spectrum showing the O $K$ edge and Os $M$ edge. (J–M) EELS mapping of $OsO_2$: (J) HAADF image; (K) Os map; (L) O map; (M) merged Os and O map. (N) Survey XPS spectrum of $OsO_2$ showing the characteristic peaks of O $1s$, C $1s$, and Os $4f$. (O) High-resolution XPS spectrum with $Os^{4+}$ $4f_{7/2}$, $4f_{5/2}$, and $5p_{3/2}$ peaks. (P) High-resolution XPS spectrum of the O $1s$ region with lattice oxygen and oxygen-related species peaks. (Q) Raman spectrum of $OsO_2$ showing vibrational modes assigned to $E_g$ (544 cm$^{-1}$), $A_{1g}$ (683 cm$^{-1}$), and $B_{2g}$ (725 cm$^{-1}$).

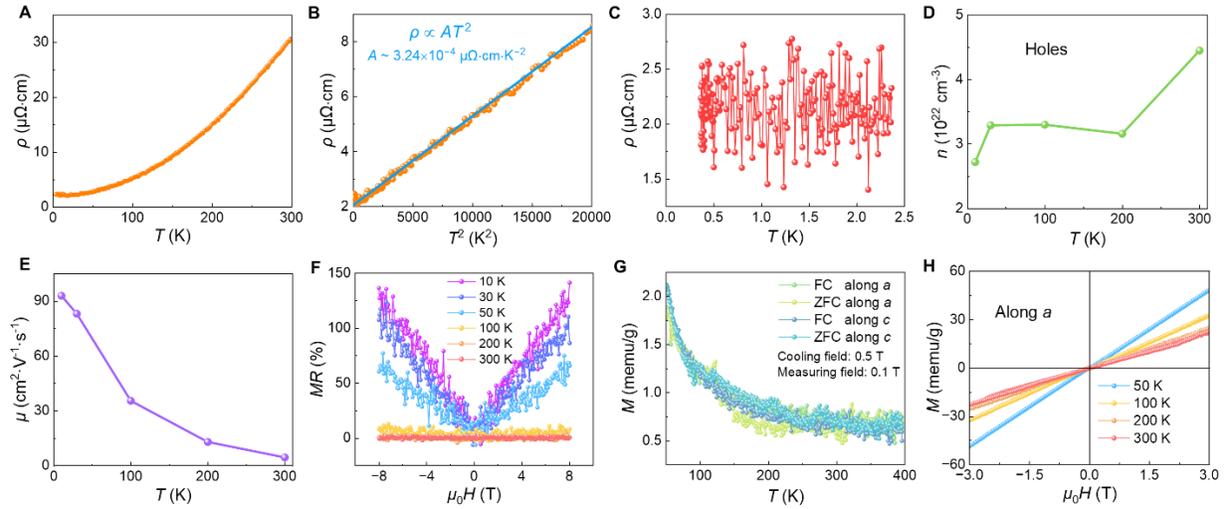

**Figure 2. Electrical and magnetic transport properties of the OsO$_2$ single crystal.**
(A) Temperature dependence of resistivity ($\rho$) of OsO$_2$ single crystal from 2 K to 300 K. (B) Low-temperature resistivity plotted as a function of $T^2$, showing a linear relationship consistent with Fermi liquid behavior. (C) Resistivity versus temperature between 2.3 and 0.37 K measured by the Helium-3 module. (D) Temperature-dependent carrier concentration ($n$). (E) Mobility ($\mu$) as a function of temperature. (F) Magnetoresistance ($MR$) curves measured under various temperatures (10, 30, 50, 100, 200 and 300 K). (G) Temperature-dependent magnetization $M(T)$ measured under zero-field-cooled (ZFC) and field-cooled (FC) conditions along the $a$- and $c$-axes. A cooling field of 0.5 T and a measuring field of 0.1 T were applied. (H) Field-dependent magnetization $M(\mu_0 H)$ curves measured along the $a$-axis at various temperatures.

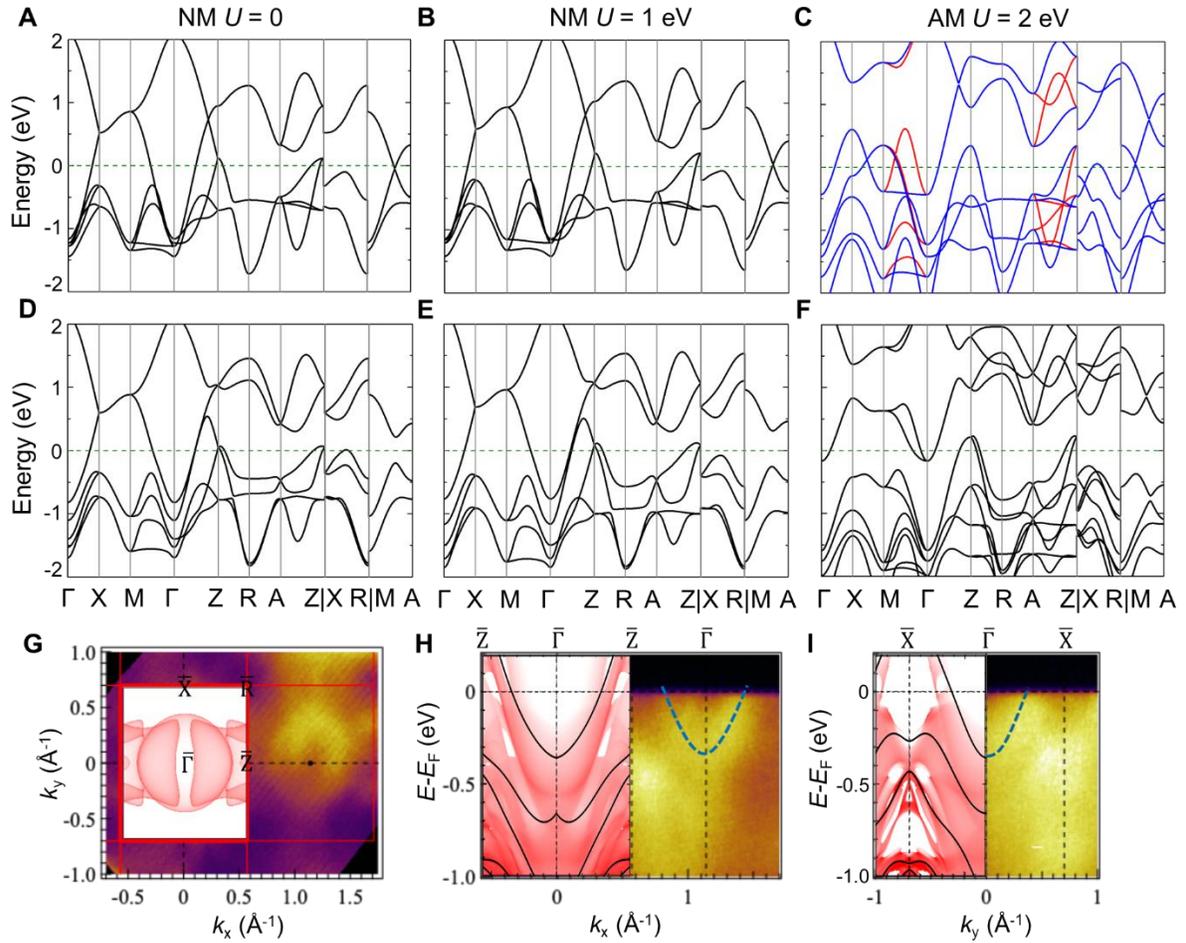

**Figure 3. Band structure and ARPES characterization of an OsO$_2$ single crystal.**
(A-C) Calculated band structures without SOC using on-site Coulomb correlation $U$ of 0, 1 and 2 eV, respectively. The former two give a nonmagnetic (NM) ground state, while the $U$ = 2 eV results in an altermagnetic (AM) ground state. (D-F) Calculated band structures with SOC using on-site Coulomb correlation $U$ of 0, 1 and 2 eV, respectively. (G) ARPES measured Fermi surface of OsO$_2$ (101) plane, using photon energy 100 eV. Superposed in the first BZ is the DFT simulated Fermi surface projected to (101) surface. (H, I) ARPES measured band spectra along high symmetry paths. Calculated band projections onto the (101) surface are also superposed.

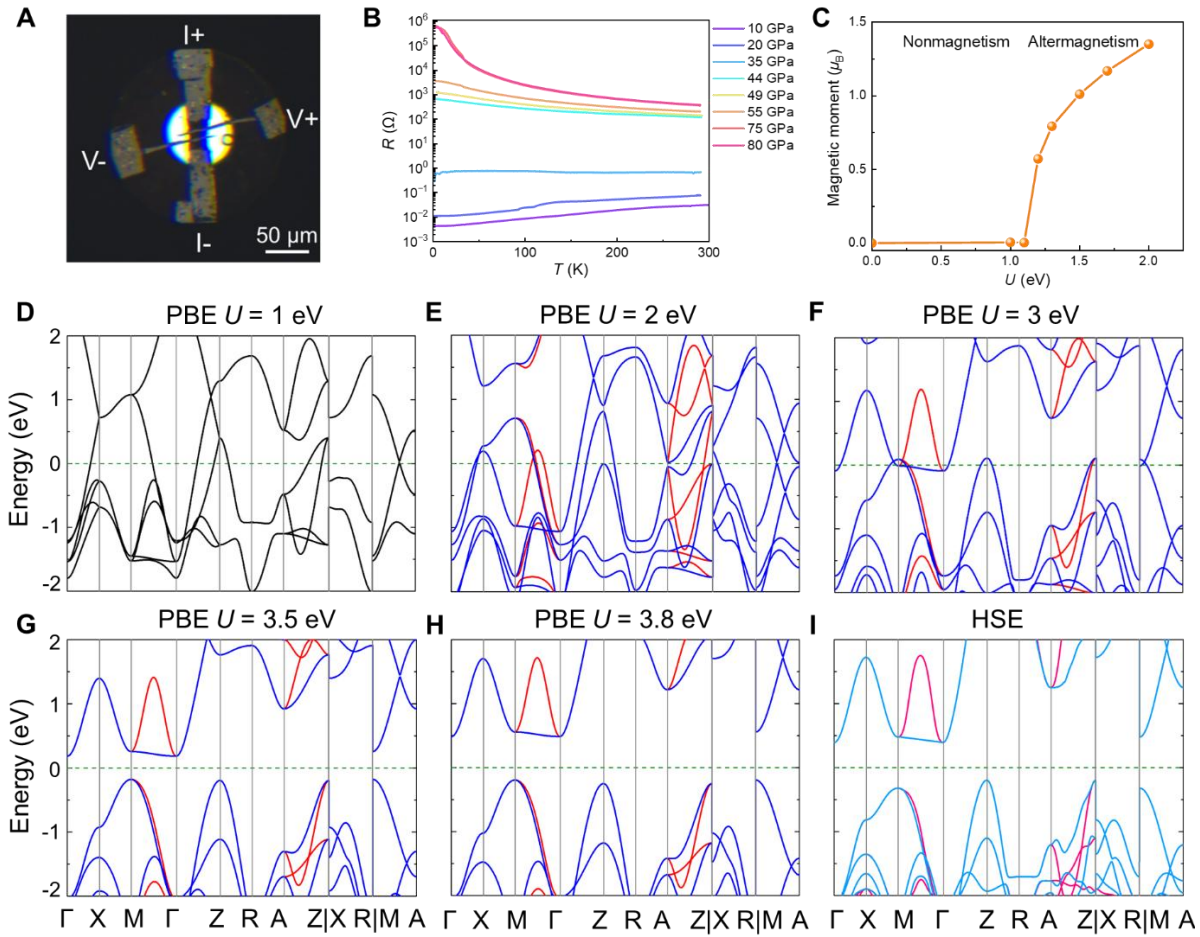

**Figure 4. High-pressure electrical transport and correlation-driven magnetic transition of an $OsO_2$ single crystal.**
(A) Optical image of a several-micrometer-thick $OsO_2$ single-crystal flake with linear four-probe electrical contacts inside a diamond anvil cell (DAC). (B) Temperature-dependent resistance of the $OsO_2$ flake under various DAC pressures ranging from 10 to 80 GPa. (C) Calculated magnetic moment of bulk $OsO_2$ as a function of on-site Coulomb correlation $U$. The magnetic ground states are marked by different regions. (D-H) Calculated electronic band structures of bulk $OsO_2$ at 50 GPa using the PBE+$U$ method with $U$ = 1, 2, 3, 3.5 and 3.8 eV, respectively. (I) Calculated electronic band structure of bulk $OsO_2$ at 50 GPa using the HSE hybrid functional. The black lines in (D) indicate nonmagnetic state, while red/blue lines in (E-I) indicate altermagnetic state.

**REFERENCES**

1. Mazin, I. I. (2022). Editorial: altermagnetism—a new punch line of fundamental magnetism, Phys. Rev. X *12*, 040002.

2. Šmejkal, L., Sinova, J., Jungwirth, T. (2022). Emerging research landscape of altermagnetism, Phys. Rev. X *12*, 040501.

3. Šmejkal, L., Sinova, J., Jungwirth, T. (2022). Beyond conventional ferromagnetism and antiferromagnetism: a phase with nonrelativistic spin and crystal rotation symmetry, Phys. Rev. X *12*, 031042.

4. Šmejkal, L., González-Hernández, R., Jungwirth, T., Sinova, J. (2020). Crystal time-reversal symmetry breaking and spontaneous Hall effect in collinear antiferromagnets, Sci. Adv. *6*, eaaz8809.

5. Berlijn, T., Snijders, P. C., Delaire, O., Zhou, H.-D., Maier, T. A., Cao, H.-B., Chi, S.-X., Matsuda, M., Wang, Y., Koehler, M. R., et al. (2017). Itinerant antiferromagnetism in $RuO_2$, Phys. Rev. Lett. *118*, 077201.

6. Zhu, Z. H., Strempfer, J., Rao, R. R., Occhialini, C. A., Pelliciari, J., Choi, Y., Kawaguchi, T., You, H., Mitchell, J. F., Shao-Horn, Y., et al. (2019). Anomalous antiferromagnetism in metallic $RuO_2$ determined by resonant X-ray scattering, Phys. Rev. Lett. *122*, 017202.

7. Keßler, P., Garcia-Gassull, L., Suter, A., Prokscha, T., Salman, Z., Khalyavin, D., Manuel, P., Orlandi, F., Mazin, I. I., Valentí, R., et al. (2024). Absence of magnetic order in $RuO_2$: insights from µSR spectroscopy and neutron diffraction, npj Spintronics *2*, 50.

8. Kiefer, L., Wirth, F., Bertin, A., Becker, P., Bohatý, L., Schmalzl, K., Stunault, A., Rodríguez-Velamazan, J. A., Fabelo, O., Braden, M. (2025). Crystal structure and absence of magnetic order in single-crystalline $RuO_2$, J. Phys.: Condens. Matter *37*, 135801.

9. Liu, J., Zhan, J., Li, T., Liu, J., Cheng, S., Shi, Y., Deng, L., Zhang, M., Li, C., Ding, J., et al. (2024). Absence of altermagnetic spin splitting character in rutile oxide $RuO_2$, Phys. Rev. Lett. *133*, 176401.

10. Osumi, T., Yamauchi, K., Souma, S., Shubhankar, P., Honma, A., Nakayama, K., Ozawa, K., Kitamura, M., Horiba, K., Kumigashira, H., et al. (2025). Spin-degenerate bulk bands and topological surface states of $RuO_2$, arXiv:2501.10649.

11. Hiraishi, M., Okabe, H., Koda, A., Kadono, R., Muroi, T., Hirai, D., Hiroi, Z. (2024). Nonmagnetic ground state in $RuO_2$ revealed by muon spin rotation, Phys. Rev. Lett. *132*, 166702.

12. Wenzel, M., Uykur, E., Rößler, S., Schmidt, M., Janson, O., Tiwari, A., Dressel, M., Tsirlin, A. A. (2025). Fermi-liquid behavior of nonaltermagnetic $RuO_2$, Phys. Rev. B *111*, L041115.

13. Pawula, F., Fakih, A., Daou, R., Hébert, S., Mordvinova, N., Lebedev, O., Pelloquin, D., Maignan, A. (2024). Multiband transport in $RuO_2$, Phys. Rev. B *110*, 064432.

14. Wang, Z. Q., Li, Z. Q., Sun, L., Zhang, Z. Y., He, K., Niu, H., Cheng, J., Yang, M., Yang, X., Chen, G., et al. (2024). Inverse spin Hall effect dominated spin-charge conversion in (101) and (110)-oriented $RuO_2$ films, Phys. Rev. Lett. *133*, 046701.

15. Peng, X., Liu, Z., Zhang, S., Zhou, Y., Sun, Y., Su, Y., Wu, C., Zhou, T., Liu, L., Li, Y., Wang, H., et al. (2025). Universal scaling behavior of transport properties in non-magnetic $RuO_2$, Commun. Mater. *6*, 177.

16. Feng, Z., Zhou, X., Šmejkal, L., Wu, L., Zhu, Z., Guo, H., González-Hernández, R., Wang, X., Yan, H., Qin, P., et al. (2022). An anomalous Hall effect in altermagnetic ruthenium dioxide, Nat. Electron. *5*, 735–743.

17. Tschirner, T., Keßler, P., Gonzalez Betancourt, R. D., Kotte, T., Kriegner, D., Büchner, B., Dufouleur, J., Kamp, M., Jovic, V., Smejkal, L., et al. (2023). Saturation of the anomalous Hall effect at high magnetic fields in altermagnetic $RuO_2$, APL Mater. *11*, 101103.

18. Jeong, S. G., Lee, S., Lin, B., Yang, Z., Choi, I. H., Oh, J. Y., Song, S., Lee, S. W., Nair, S., Choudhary, R., et al. (2025). Metallicity and anomalous Hall effect in epitaxially-strained, atomically-thin $RuO_2$ films, Proc. Natl. Acad. Sci. U.S.A. *122*, e2500831122.

19. Bose, A., Schreiber, N. J., Jain, R., Shao, D.-F., Nair, H. P., Sun, J., Zhang, X. S., Muller, D. A., Tsymbal, E. Y., Schlom, D. G., et al. (2022). Tilted spin current generated by the collinear antiferromagnet ruthenium dioxide, Nat. Electron. *5*, 267–274.

20. Wang, M., Tanaka, K., Sakai, S., Wang, Z., Deng, K., Lyu, Y., Li, C., Tian, D., Shen, S., Ogawa, N., et al. (2023). Emergent zero-field anomalous Hall effect in a reconstructed rutile antiferromagnetic metal, Nat. Commun. *14*, 8240.

21. Bai, H., Han, L., Feng, X. Y., Zhou, Y. J., Su, R. X., Wang, Q., Liao, L. Y., Zhu, W. X., Chen, X. Z., Pan, F., et al. (2022). Observation of spin splitting torque in a collinear antiferromagnet $RuO_2$, Phys. Rev. Lett. *128*, 197202.

22. Karube, S., Tanaka, T., Sugawara, D., Kadoguchi, N., Kohda, M., Nitta, J. (2022). Observation of spin-splitter torque in collinear antiferromagnetic $RuO_2$, Phys. Rev. Lett. *129*, 137201.

23. Fedchenko, O., Minár, J., Akashdeep, A., D'Souza, S. W., Vasilyev, D., Tkach, O., Odenbreit, L., Nguyen, Q., Kutnyakhov, D., Wind, N., et al. (2024). Observation of time-reversal symmetry breaking in the band structure of altermagnetic $RuO_2$, Sci. Adv. *10*, eadj4883.

24. Jeong, S. G., Choi, I. H., Nair, S., Buiarelli, L., Pourbahari, B., Young, J., Bassim, N., Hirai, D., Seo, A., Choi, W. S., et al. (2024). Altermagnetic polar metallic phase in ultra-thin epitaxially-strained $RuO_2$ films, arXiv:2405.05838.

25. Weber, M., Wust, S., Haag, L., Akashdeep, A., Leckron, K., Schmitt, C., Ramos, R., Kikkawa, T., Saitoh, E., Kläui, M., et al. (2024). All optical excitation of spin polarization in *d*-wave altermagnets, arXiv:2408.05187.

26. Guo, Y., Zhang, J., Zhu, Z., Jiang, Y., Jiang, L., Wu, C., Dong, J., Xu, X., He, W., He, B., et al. (2024). Direct and inverse spin splitting effects in altermagnetic $RuO_2$, Adv. Sci. *11*, 2400967.

27. Liu, Y., Bai, H., Song, Y., Ji, Z., Lou, S., Zhang, Z., Song, C., Jin, Q. (2023). Inverse altermagnetic spin splitting effect-induced terahertz emission in $RuO_2$, Adv. Opt. Mater. *11*, 2300177.

28. Zhang, Y.-C., Bai, H., Zhang, D.-H., Chen, C., Han, L., Liang, S.-X., Chu, R.-Y., Dai, J.-K., Sawicki, M., Pan, F., et al. (2025). Probing the Néel order in altermagnetic $RuO_2$ films via X-ray magnetic linear dichroism, Chin. Phys. Lett. *42*, 027301.

29. Noh, S., Kim, G.-H., Lee, J., Jung, H., Seo, U., So, G., Lee, J., Lee, S., Park, M., Yang, S., et al. (2025). Tunneling magnetoresistance in altermagnetic $RuO_2$-based magnetic tunnel junctions, Phys. Rev. Lett. *134*, 246703.

30. Chen, H., Wang, Z., Qin, P., Meng, Z., Zhou, X., Wang, X., Liu, L., Zhao, G., Duan, Z., Zhang, T., et al. (2025). Spin-splitting magnetoresistance in altermagnetic $RuO_2$ thin films, Adv. Mater. *37*, 2507764.

31. Over, H. (2012). Surface chemistry of ruthenium dioxide in heterogeneous catalysis and electrocatalysis: from fundamental to applied research, Chem. Rev. *112*, 3356–3426.

32. Rogers, D. B., Shannon, R. D., Sleight, A. W., Gillson, J. L. (1969). Crystal chemistry of metal dioxides with rutile-related structures, Inorg. Chem. *8*, 841–849.

33. Yen, P. C., Chen, R. S., Chen, C. C., Huang, Y. S., Tiong, K. K. (2004). Growth and characterization of $OsO_2$ single crystals, J. Cryst. Growth *262*, 271–276.

34. Raghuvanshi, P. R., Berlijn, T., Parker, D. S., Wang, S., Manley, M. E., Hermann, R. P., Lindsay, L., Cooper, V. R. (2025). Altermagnetic behavior in $OsO_2$: parallels with $RuO_2$, Phys. Rev. Mater. *9*, 034407.

35. Zhang, W., Zeng, M., Liu, Y., Zhang, Z., Xiong, R., Lu, Z., (2025). Strain-induced nonrelativistic altermagnetic spin splitting effect, Phys. Rev. B *112*, 024415.

36. Regoutz, A., Ganose, A. M., Blumenthal, L., Schlueter, C., Lee, T.-L., Kieslich, G., Cheetham, A. K., Kerherve, G., Huang, Y.-S., Chen, R.-S., et al. (2019). Insights into the electronic structure of $OsO_2$ using soft and hard x-ray photoelectron spectroscopy in combination with density functional theory, Phys. Rev. Mater. *3*, 025001

37. Yen, P. C., Chen, R. S., Huang, Y. S., Chia, C. T., Chen, R. H., Tiong, K. K. (2003). The first-order Raman spectra of $OsO_2$, J. Phys.: Condens. Matter *15*, 1487–1494.

38. Ryden, W. D., Lawson, A. W., Sartain, C. C. (1968). Temperature dependence of the resistivity of $RuO_2$ and $IrO_2$, Phys. Lett. A *26*, 209–210.

39. v. Löhneysen, H., Rosch, A., Vojta, M., Wölfle, P. (2007). Fermi-liquid instabilities at magnetic quantum phase transitions, Rev. Mod. Phys. *79*, 1015–1087.

40. Liu, Z. Q., Lü, W. M., Wang, X., Huang, Z., Annadi, A., Zeng, S. W., Venkatesan, T., Ariando (2012). Magnetic-field induced resistivity minimum with in-plane linear magnetoresistance of the Fermi liquid in $SrTiO_{3-x}$ single crystals, Phys. Rev. B *85*, 155114.

41. Seng, P., Diehl, J., Klimm, S., Horn, S., Tidecks, R., Samwer, K., Hänsel, H., Gross, R. (1995). Hall effect and magnetoresistance in $Nd_{1.85}Ce_{0.15}CuO_{4-\delta}$ films, Phys. Rev. B *52*, 3071–3074.

42. Maeno, Y., Yoshida, K., Hashimoto, H., Nishizaki, S., Ikeda, S., Nohara, M., Fujita, T., Mackenzie, A. P., Hussey, N. E., Bednorz, J. G., et al. (1997). Two-dimensional Fermi liquid behavior of the superconductor $Sr_2RuO_4$, J. Phys. Soc. Jpn. *66*, 1405–1408.

43. Schriempf, J. T., Schindler, A. I., Mills, D. L. (1969). Effect of electron-electron scattering on the low-temperature Lorenz numbers of dilute Pd-Ni Alloys, Phys. Rev. *187*, 959–973.

44. Paul, S., Ikeda, A., Matsuki, H., Mattoni, G., Schmalian, J., Sow, C., Yonezawa, S., Maeno, Y., (2025). Nonanalytic Fermi-liquid correction to the specific heat of $RuO_2$, arXiv: 2512.03108.

45. Andres, K., Graebner, J. E., Ott, H. R. (1975). 4*f*-virtual-bound-state formation in $CeAl_3$ at low temperatures, Phys. Rev. Lett. *35*, 1779–1782.

46. Liu, Z. Q., Leusink, D. P., Wang, X., Lü, W. M., Gopinadhan, K., Annadi, A., Zhao, Y. L., Huang, X. H., Zeng, S. W., Huang, Z., et al. (2011). Metal-insulator transition in $SrTiO_{3-x}$ thin films induced by frozen-out carriers, Phys. Rev. Lett. *107*, 146802.

47. Liu, Z. Q., Li, C. J., Lü, W. M., Huang, X. H., Huang, Z., Zeng, S. W., Qiu, X. P., Huang, L. S., Annadi, A., Chen, J. S., et al. (2013). Origin of the two-dimensional electron gas at $LaAlO_3/SrTiO_3$ interfaces: the role of oxygen vacancies and electronic reconstruction, Phys. Rev. X *3*, 021010.

48. Feng, Z., Qin, P., Yang, Y., Yan, H., Guo, H., Wang, X., Zhou, X., Han, Y., Yi, J., Qi, D., et al. (2021). A two-dimensional electron gas based on a 5*s* oxide with high room-temperature mobility and strain sensitivity, Acta Mater. *204*, 116516.

49. Son, J., Moetakef, P., Jalan, B., Bierwagen, O., Wright, N. J., Engel-Herbert, R., Stemmer, S. (2010). Epitaxial $SrTiO_3$ films with electron mobilities exceeding 30,000 $cm^2\,V^{-1}\,s^{-1}$, Nat. Mater. *9*, 482–484.

50. Yue, J., Ayino, Y., Truttmann, T. K., Gastiasoro, M. N., Persky, E., Khanukov, A., Lee, D., Thoutam, L. R., Kalisky, B., Fernandes, R. M., Pribiag, V. S., Jalan, B. (2022). Anomalous transport in high-mobility superconducting $SrTiO_3$ thin films, Sci. Adv. *8*, eabl5668.

51. Hu, J., Rosenbaum, T. F. (2008). Classical and quantum routes to linear magnetoresistance, Nat. Mater. *7*, 697.

52. Xu, R., Husmann, A., Rosenbaum, T. F., Saboungi, M.-L., Enderby, J. E., Littlewood, P. B. (1997). Large magnetoresistance in non-magnetic silver chalcogenides, Nature *390*, 57–60.

53. Yang, J., Song, Z.-Y., Guo, L., Gao, H., Dong, Z., Yu, Q., Zheng, R.-K., Kang, T.-T., Zhang, K. (2021). Nontrivial giant linear magnetoresistance in nodal-line semimetal ZrGeSe 2D layers, Nano Lett. *21*, 10139–10145.

54. Feng, Y., Wang, Y., Silevitch, D. M., Yan, J.-Q., Kobayashi, R., Hedo, M., Nakama, T., Ōnuki, Y., Suslov, A. V., Mihaila, B., et al. (2019). Linear magnetoresistance in the low-field limit in density-wave materials, Proc. Natl. Acad. Sci. U.S.A. *116*, 11201–11206.

55. Zhou, X., Liu, Z. (2021). Nature of electrons from oxygen vacancies and polar catastrophe at LaAlO$_3$/SrTiO$_3$ interfaces, J. Phys.: Condens. Matter *33*, 435601.

56. Liu, Z., Chen, H., Wang, J., Liu, J., Wang, K., Feng, Z., Yan, H., Wang, X., Jiang, C., Coey, J. M. D., et al. (2018). Electrical switching of the topological anomalous Hall effect in a non-collinear antiferromagnet above room temperature, Nat. Electron. *1*, 172–177.

57. Yan, H., Feng, Z. X., Shang, S., Wang, X., Hu, Z., Wang, J., Zhu, Z., Chen, Z., Hua, H., Lu, W., et al. (2019). A piezoelectric, strain-controlled antiferromagnetic memory insensitive to magnetic fields, Nat. Nanotechnol. *14*, 131–136.

58. Sun, H., Huo, M., Hu, X., Li, J., Liu, Z., Han, Y., Tang, L., Mao, Z., Yang, P., Wang, B., et al. (2023). Signatures of superconductivity near 80 K in a nickelate under high pressure, Nature *621*, 493–498.

59. Yan, H., Feng, Z. X., Qin, P., Zhou, X., Guo, H., Wang, X., Chen, H., Zhang, X., Wu, H., Jiang, C., et al. (2020). Electric-field-controlled antiferromagnetic spintronic devices, Adv. Mater. *32*, 1905603.

60. Guo, H., Feng, Z., Yan, H., Liu, J., Zhang, J., Zhou, X., Qin, P., Cai, J. L., Zeng, Z., Zhang, X., et al. (2020). Giant piezospintronic effect in a noncollinear antiferromagnetic metal, Adv. Mater. *32*, 2002300.

61. Qin, P., Feng, Z., Zhou, X., Guo, H., Wang, J., Yan, H., Wang, X., Chen, H., Zhang, X., Wu, H., et al. (2020). Anomalous Hall effect, robust negative magnetoresistance, and memory devices based on a noncollinear antiferromagnetic metal, ACS Nano *14*, 6242–6248.

62. Chen, H., Feng, Z., Qin, P., Zhou, X., Yan, H., Wang, X., Meng, Z., Liu, L., Liu, Z. (2022). Resistive memory based on the spin-density-wave transition of antiferromagnetic chromium, Phys. Rev. Appl. *18*, 054046.

63. Chen, H., Zhou, X., Meng, Z., Wang, X., Duan, H., Liu, L., Zhao, G., Yan, H., Qin, P., Liu, Z. (2024). Magnetic-field response and giant electric-field modulation of Cu$_2$S, Nano Lett. *24*, 584–591.

64. Wang, X., Feng, Z., Qin, P., Yan, H., Zhou, X., Guo, H., Leng, Z., Chen, W., Jia, Q., Hu, Z., et al. (2019). Integration of the noncollinear antiferromagnetic metal Mn$_3$Sn onto ferroelectric oxides for electric-field control, Acta Mater. *181*, 537–543.

65. Kresse, G., Furthmuller, J. (1996). Efficient iterative schemes for ab initio total-energy calculations using a plane-wave basis set. Phys. Rev. B *54*, 11169.

66. Blöchl, P. E. (1994). Projector augmented-wave method. Phys. Rev. B *50*, 17953.

67. Perdew, J. P., Burke, K., Ernzerhof, M. (1996). Generalized Gradient Approximation Made Simple. Phys. Rev. Lett. *77*, 3865.

68. Dudarev, S. L., Botton, G. A., Savrasov, S. Y., Humphreys, C. J., Sutton, A. P. (1998). Electron-energy-loss spectra and the structural stability of nickel oxide: An LSDA+U study. Phys. Rev. B *57*, 1505.

69. Paier, J., Marsman, M., Hummer, K., Kresse, G., Gerber, I. C., Ángyán, J. G. (2006). Screened hybrid density functionals applied to solids. J. Chem. Phys. *124*, 154709.

70. Souza, I., Marzari, N., Vanderbilt, D. (2001). Maximally localized Wannier functions for entangled energy bands. Phys. Rev. B *65*, 035109.

71. Wu, Q., Zhang, S., Song, H.-F., Troyer, M., Soluyanov, A. A. (2018). WannierTools: An open-source software package for novel topological materials. Comput. Phys. Commun. *224*, 405.

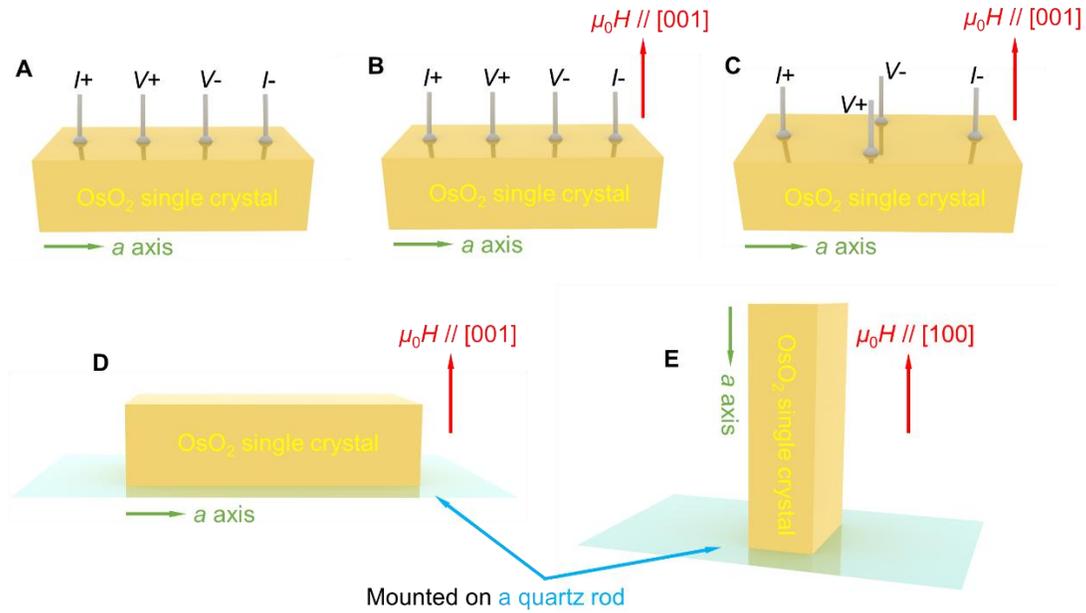

**Fig. S1. Measurement geometries for transport and magnetic measurements on $OsO_2$ single crystals.** (A) Four-probe configuration for electrical transport with the current applied along the crystallographic *a* axis. (B) Magnetoresistance measurement geometry with the current along the *a* axis and the magnetic field applied perpendicular to the sample surface. (C) Hall measurement geometry with the current along the *a* axis, transverse Hall voltage measured within the sample plane, and the magnetic field applied along [001]. (D) Out-of-plane magnetization measurement with the magnetic field applied perpendicular to the sample surface. (E) In-plane magnetization measurement with the magnetic field applied parallel to the [100] direction.

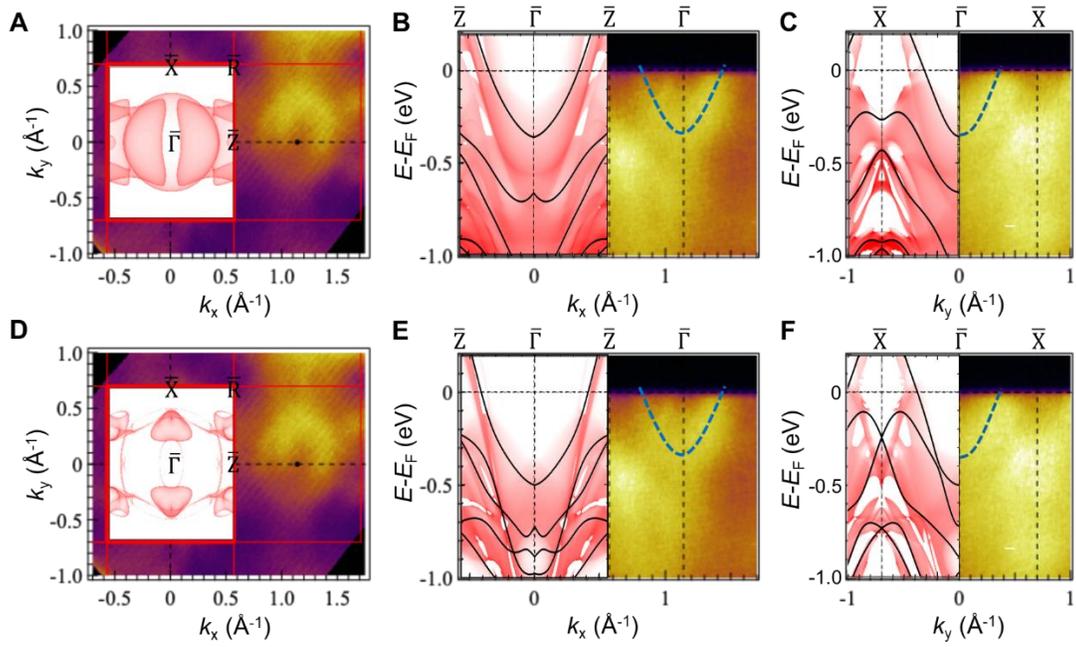

**Fig. S2. Comparison between ARPES measurements and DFT calculations with and without SOC.** Comparison between ARPES data and DFT calculations with (A–C) and without (D–F) SOC by using an on-site Coulomb interaction of $U$ = 1 eV. The red shaded areas represent the projection of bulk bands onto the (101) surface, while the black lines correspond to the bulk bands at $k_z = 0.25\pi/c'$, where $c'$ is the lattice vector perpendicular to the (101) direction.

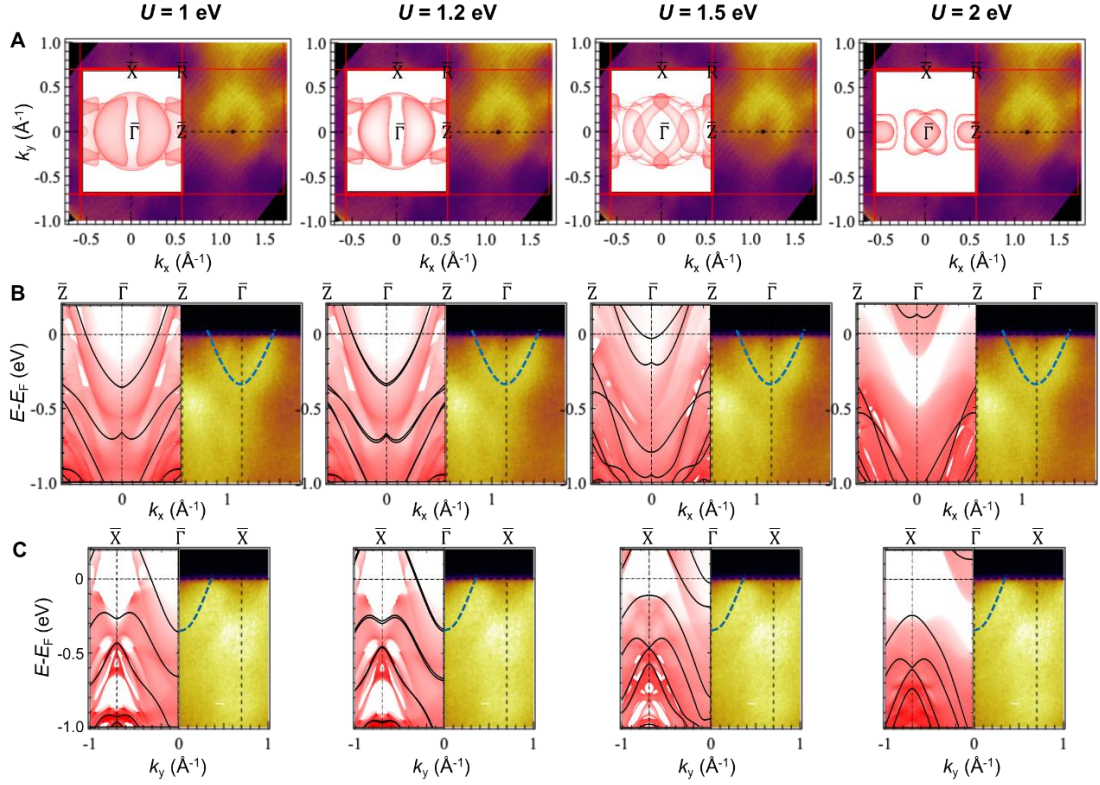

**Fig. S3. ARPES–DFT comparison with varying on-site Coulomb interaction U.** Comparison between ARPES data and DFT calculations with different $U$ for Fermi surface (A), band structures along $\bar{Z} - \bar{\Gamma} - \bar{Z}$ (B) and along $\bar{\Gamma} - \bar{X} - \bar{\Gamma}$ (C). The SOC effect is considered in the calculations. The red shaded areas represent the projection of bulk bands onto the (101) surface, while the black lines correspond to the bulk bands at $k_z = 0.25\pi/c'$, where $c'$ is the lattice vector perpendicular to the (101) direction.